\documentclass[rmp,aps,floatfix,showpacs,showkeys]{revtex4-1}
\usepackage{graphicx} 
\usepackage{epstopdf}
\usepackage{dcolumn}
\usepackage{bm}
\usepackage{amsfonts,amsmath,amssymb
}

\usepackage{bm,dsfont}

\usepackage{bbm}
\usepackage{longtable}
\usepackage[dvips]{epsfig}
\usepackage{hyperref}
\usepackage{listings}
\usepackage{color}
\DeclareMathOperator{\Tr}{Tr}

\newcommand{\unit}{1\!\!1}

\let\phi\varphi

\let\theta\vartheta
\renewcommand{\rho}{\varrho}

\newcommand{\tr}[1]{\operatorname{Tr} #1}

\begin{document}
\title{Conditions for the existence of
positive operator valued measures}
\author{Maximilian Schumacher}
\author{Gernot Alber}
\affiliation{Institut f\"{u}r Angewandte Physik, Technische 
Universit\"at Darmstadt, D-64289 Darmstadt, Germany}
\date{\today}
\begin{abstract}
{Sufficient and necessary conditions are presented for the existence of
$(N,M)$-positive operator valued measures ($(N,M)$-POVMs) valid for arbitrary-dimensional quantum systems.
A sufficient condition for the existence of $(N,M)$-POVMs is presented.
It yields a simple relation determining an upper bound on the continuous parameter of an arbitrary $(N,M)$-POVM, below which all its POVM elements are guaranteed to be positive semidefinite. Necessary conditions are derived for the existence of  optimal $(N,M)$-POVMs. One of these necessary conditions exhibits a close connection between the existence of optimal informationally complete $(N,M)$-POVMs and the existence of isospectral, traceless, orthonormal, hermitian operator bases in cases, in which the parameter $M$ exceeds the dimension of the quantum system under consideration.
Another necessary condition is derived for optimal $(N,M)$-POVMs, whose parameter $M$ is less than the dimension of the quantum system.
It is shown that in these latter cases all POVM elements necessarily are projection operators of equal rank. This significantly constrains the possible parameters for constructing optimal $(N,M)$-POVMs. For the special case of $M=2$  a necessary and sufficient condition for the existence of optimal $(N,2)$-POVMs is presented.
}
\end{abstract}
\maketitle
\section{Introduction}
The development of efficient quantum measurement techniques is important for central tasks of quantum information processing \cite{measurement-general,POVM2}, such as quantum state  reconstruction or the detection of characteristic quantum correlations. In this context $(N,M)$-POVMs \cite{NMPOVM} have been introduced recently as interesting  one-parameter-continuous families of  positive operator valued measures (POVMs). They describe numerous important quantum measurements in a unified way.  These measurements include projective measurements with complete sets of mutually unbiased bases (MUBs) \cite{MUB}, mutually unbiased measurements \cite{MUM}, symmetric informationally complete POVMs (SIC-POVMs) \cite{SIC-POVM1,SIC-POVM2} and their generalizations (GSIC-POVMs) \cite{GSIC-POVM}. For purposes of quantum information processing informationally complete $(N,M)$-POVMs are particularly interesting, because they enable a complete reconstruction of quantum states. 

Recent investigations exploring characteristic features of $(N,M)$-POVMs have concentrated
on possible applications in quantum information processing  and on basic theoretical questions concerning their existence.
On the application side the potential of $(N,M)$-POVMs for the local detection of provable bipartite quantum entanglement \cite{Schumacher-Alber-PRA} and quantum steering \cite{Schumacher-Alber-PhysScr} has been investigated. 
As far as  basic theoretical questions are concerned, it has been demonstrated that $(N,M)$-POVMs can always be constructed for sufficiently small values of their continuous parameter \cite{NMPOVM}. However, for larger or even maximal values of their continuous parameter, i.e. for optimal $(N,M)$-POVMs, it generally causes major theoretical problems to guarantee the positive semidefiniteness of all POVM elements involved.
Despite considerable research efforts concentrating on the subclasses of SIC-POVM and MUBs, for example, open questions concerning their existence in arbitrary dimensions still remain \cite{FIVE-OPEN,GSIC-construction}.
Thus, questions concerning the existence and construction of $(N,M)$-POVMs for large or even maximal values of their continuous  parameter are still widely open. 

In order to obtain a detailed theoretical understanding of $(N,M)$-POVMs and of their characteristic features there is a need to develop sufficient and necessary conditions, which guarantee their existence. It is a main intention of this paper to address this issue
 and to explore general features of the existence and construction of $(N,M)$-POVMs with the help of orthonormal hermitian operator bases. 
As a first main result we develop a sufficient condition for the existence of  $(N,M)$-POVMs. This sufficient condition yields a simple upper bound on the continuous parameter, within which  this existence is guaranteed. Thus, this result complements the already known property that $(N,M)$-POVMs can always be constructed for sufficiently small values of their continuous parameters \cite{NMPOVM}.
As a second main result we present necessary conditions for the existence of optimal $(N,M)$-POVMs. One of these necessary conditions exhibits  close connections between the existence of optimal informationally complete $(N,M)$-POVMs and the existence of isospectral, traceless, orthonormal, hermitian operator bases in cases, in which the parameter $M$ exceeds the dimension of the quantum system to be measured. This result is based on recent work showing that informationally complete $(N,M)$-POVMs are necessarily related to orthonormal hermitian operator bases by highly degenerate linear maps with two non-zero eigenvalues and a non-trivial kernel \cite{Schumacher-Alber-PRA}.  Thus, it generalizes an already known property of GSICs \cite{GSIC-POVM}, i.e. optimal $(1,M)$-POVMs, to arbitrary optimal informationally complete $(N,M)$-POVMs.
Another necessary condition shows that optimal $(N,M)$-POVMs, whose parameter $M$ is less than the dimension of the quantum system under consideration, can necessarily only exist if all their POVM elements are projection operators of equal rank. This significantly constrains the possible parameters, for which optimal $(N,M)$-POVMs can be constructed. 
Furthermore, in the special cases with $M=2$ we present a necessary and sufficient condition for the existence of optimal $(N,2)$-POVMS. This criterion establishes a connection to the existence of $N$  isospectral, traceless, orthonormal, hermitian operators in even-dimensional Hilbert spaces with a prescribed spectrum.

 This paper is organized as follows.
In Sec.\ref{Sec-1} basic features of $(N,M)$-POVMs are summarized.  
In Sec.\ref{basic}
their defining properties are recapitulated \cite{NMPOVM}.
In Sec.\ref{Sec-20} recent results \cite{Schumacher-Alber-PRA} on their necessary relation to orthonormal hermitian operator bases are summarized. Furthermore, as a motivation for the subsequent discussion typical problems are discussed originating from the positive semidefiniteness of all its POVM elements.
As one of our main results, in Sec.\ref{SecIII}
a sufficient condition is presented, under which  for any $d$-dimensional quantum system $(N,M)$-POVMs can be constructed.  As a second main result
in Sec.\ref{Sec-necc}
two necessary conditions for the existence of optimal  $(N,M)$-POVMs are presented. Firstly, it is shown that for $M\geq d$ the existence of $(d^2-1)$ isospectral,  traceless, orthonormal, hermitian operators is necessary for the existence of an optimal informationally complete $(N,M)$-POVM. Thereby the form of their common spectrum  is completely determined by the defining properties of the $(N,M)$-POVM. 
Secondly, it is shown that in cases with $2 < M <d$
optimal $(N,M)$-POVMs can necessarily only exist for equal-rank projection operators. 
In Sec.\ref{Sec-3} a necessary and sufficient condition for the existence of optimal $(N,2)$-POVMs with $N\leq d^2-1$ is derived.
It is shown that they can only exist in even-dimensional Hilbert spaces. Explicit constructions are presented for dimensions $d=2^k$ with $k\in  {\mathbb N}$. 

\section{Informationally complete positive operator valued measures\label{Sec-1}}
In this section elementary features of the recently developed (N,M)-POVMs \cite{NMPOVM} are summarized.
In the first subsection their definition and resulting elementary properties are recapitulated. In the second subsection recently discussed basic relations between informationally complete $(N,M)$-POVMs and orthonormal hermitian operator bases are summarized \cite{Schumacher-Alber-PRA}. Furthermore, some typical problems are discussed, which complicate the construction of $(N,M)$-POVMs by basis expansions in terms of orthonormal hermitian operator bases.

\subsection{Basic properties\label{basic}}

Let us consider a quantum system characterized by a $d$-dimensional Hilbert space. Ignoring the properties of a quantum state immediately after a measurement, the most general quantum measurement on this quantum system is described by a POVM \cite{measurement-general,POVM2}.
A $M$-element POVM is a set of $M$ positive semidefinite operators, say $\Pi = \{\Pi_a \geq 0 \mid a=1,\cdots,M\}$, which fulfill the completeness relation
\begin{eqnarray}
\sum_{a=1}^M \Pi_a &=&\unit_d
\label{complete1}
\end{eqnarray}
with the unit operator $\unit_d$ of the quantum system's $d$-dimensional Hilbert space.
Thereby, the indices $a \in \{1,\cdots,M\}$ coordinatize the different $M$ possible real-valued measurement results, say ${\cal M}_a$.
According to Born's rule the probability of measuring the result ${\cal M}_a$ is given by
$p_a = \Tr\{\rho \Pi_a\}$,  if the quantum system has been prepared in the quantum state $\rho \geq 0$ immediately before the measurement. If the positive semidefinite operators $\Pi$ are linearly independent for $a \in \{1,\cdots,M\}$ and $M= d^2$, a POVM is called informationally complete. In such a case an arbitrary quantum state $\rho$ can be reconstructed from all the $d^2$ measurement results.
In the special case of orthogonal projection operators, i.e. $\Pi_a \Pi_{a'} = \delta_{a a'}\Pi_a$ for $a,a' \in \{1,\cdots,M\}$, a POVM describes a von Neumann measurement.

Recently $(N,M)$-POVMs \cite{NMPOVM} have been introduced as a unified way for describing numerous important quantum measurements, such as projective measurements with MUBs \cite{MUB}, MUMs \cite{MUM}, SIC-POVMs \cite{SIC-POVM1,SIC-POVM2} and their generalizations (GSICs) \cite{GSIC-POVM}.
A $(N,M)$-POVM $\Pi$ is a one-continuous-parameter family of $N$ different $M$-element POVMs, i.e. $\Pi = \{\Pi_{i(\alpha,a)}\mid \alpha \in \{1,\cdots,N\}, a\in \{1,\cdots,M\}\}$, 
defined by the following relations
\begin{eqnarray}
\Tr\{\Pi_{i(\alpha,a)} \} &=&\frac{d}{M},\label{add1}\\
\Tr\{\Pi_{i(\alpha,a)} ~\Pi_{i(\alpha,a')} \} &=&x ~\delta_{a,a'} + (1-\delta_{a,a'})\frac{d-Mx}{M(M-1)},\label{add2}\\
\Tr\{\Pi_{i(\alpha,a)} ~\Pi_{j(\beta,b)} \} &=& \frac{d}{M^2}
\label{add3}
\end{eqnarray}
for all $\beta\neq \alpha \in \{1,\cdots,N\}$ and $a,a',b \in \{1,\cdots,M\}$.
Thereby, for the sake of convenience we have introduced the coordinate function $i: (\alpha,a) \to i(\alpha,a)$. It maps the $N$ $M$-tuples of the form $(\alpha,a)$,
 which
 identify a particular POVM element uniquely, 
  bijectively onto the $NM$ natural numbers  $i,j\in \{1,\cdots, NM\}$.
For given values of $(d,N,M)$ 
the possible values of the continuous parameter $x$ are constrained by the relation \cite{NMPOVM}
\begin{eqnarray}
\frac{d}{M^2}< x \leq {\rm min}\left(\frac{d^2}{M^2},\frac{d}{M}\right).
\label{maximal}
\end{eqnarray}
A $(N,M)$-POVM with maximal possible value of $x$ is called optimal. 
Furthermore, a $(N,M)$-POVM $\Pi$ is informationally complete, if it contains $d^2$ linearly independent positive semidefinite operators. As each of the $N$ $M$-element POVMs involved fulfills the completeness relation (\ref{complete1}),  this is equivalent to the requirement 
\begin{eqnarray}
(M-1)N + 1 &=& d^2.
\label{dimension5}
\end{eqnarray}
For arbitrary dimensions four possible
solutions of this relation are
 $(N,M) \in \{(1,d^2),~(d+1,d),~(d^2-1,2),~(d-1,d+2)\}$.
The solution $(N,M) = (1,d^2)$ describes a one-parameter family of
GSIC-POVMs \cite{GSIC-POVM} parameterized by the parameter $x$. SIC-POVMs are special cases of GSIC-POVMs with $x=1/d^2$.
The solution $(N,M) = (d+1,d)$ describes MUMs \cite{MUM}. In the special case of $x=d^2/M^2=d/M=1$  these MUMs describe projective measurements of unit rank with maximal sets of $d+1$ MUBs.

\subsection{Orthonormal hermitian operator bases and informationally complete (N,M)-POVMs\label{Sec-20}}
In this subsection we first recapitulate the general relations between informationally complete $(N,M)$-POVMs and orthonormal hermitian operator bases, which necessarily have to be fulfilled irrespectively of the positive semidefiniteness of the POVM elements involved \cite{Schumacher-Alber-PRA}.
They govern the construction of $(N,M)$-POVMs by basis expansions in terms of orthonormal hermitian operator bases. Subsequently,
concentrating on the particular example of a MUM in dimension $d=3$, it is exemplified that such a MUM  can always be constructed for sufficiently small values of the continuous parameter $x$. 
This example demonstrates, why constructing $(N,M)$-POVMs for parameters $x$ close to their lower bounds is always possible, while constructing optimal $(N,M)$-POVMs is rather difficult.
These problems motivate the development of simple sufficient and necessary conditions for the existence of $(N,M)$-POVMs, which is pursued in the subsequent sections.

 In a Hilbert space ${\cal H}_d$ of a $d$-dimensional quantum system an informationally complete
$(N,M)$-POVM can always be expanded in a basis of $d^2$ linearly independent linear operators, say $G=(G_1,\cdots,G_{d^2})^T$. These operators can be chosen as orthonormal hermitian operators with respect to the Hilbert-Schmidt (HS) scalar product
$\langle G_{\mu} | G_{\nu}\rangle_{HS} := \Tr\{G_{\mu}^{\dagger} G_{\nu}\}$ with $G_{\mu}^{\dagger} = G_{\mu}$. They form a basis of the Hilbert space ${\cal H}_{d^2} = ({\rm Span}(G),\langle \cdot | \cdot \rangle_{HS})$ of linear operators in ${\cal H}_d$ over the field of real numbers. This latter Hilbert space is a Euclidean vector space. Furthermore, such an orthonormal hermitian operator basis ${G}$ can always be chosen so that
\begin{eqnarray}
G_1 &=& \unit_d/\sqrt{d},~~
\Tr\{G_{\mu} \} =0 
\label{operatorbasis}
\end{eqnarray}
for $\mu \in \{2,\cdots,d^2\}$.
The resulting basis expansion of
an arbitrary informationally complete $(N,M)$-POVM in such an orthonormal hermitian basis has the general form 
\begin{eqnarray}
\Pi = G^T S.
\label{basisexpansion}
\end{eqnarray}
Thereby, $S$ denotes the linear operator mapping ${\cal H}_{d^2}$ into
the Hilbert space ${\cal H}_{NM}$ of hermitian operators, which contains all the elements of the $(N,M)$-POVM. 
Recently, it has been shown 
that for informationally complete $(N,M)$-POVMs
the structure of this linear map $S$ and of its corresponding real-valued
$d^2 \times NM$ matrix $S_{\mu,i}$ is significantly constrained by the defining relations (\ref{add2}) and  (\ref{add3}) \cite{Schumacher-Alber-PRA}.
Ignoring the positive semidefiniteness constraints of the POVM elements,
it has been shown that the most general form of the linear operator $S: {\cal H}_{d^2} \to {\cal H}_{NM}$ 
is given by a $d^2\times NM$ matrix of the form
\begin{eqnarray}
{S}_{\mu, i(\alpha,a)} &=& \sqrt{\Lambda_{\mu}}X^T_{\mu,i(\alpha,a)}
\label{S2}
\end{eqnarray}
with the diagonal $d^2\times d^2$ matrix $\Lambda$ and the $NM\times d^2$ matrix
$X_{i,\mu}$. 
The diagonal matrix $\Lambda$ has only two different non-vanishing entries, which are the non-zero eigenvalues of $S$. They are given by
\begin{eqnarray}
\Lambda_1 &=&{\frac{dN}{M}},~~
\Lambda_{\mu} = \Gamma =  \frac{x M^2 -d}{M(M-1)}
\end{eqnarray}
for $\mu \in \{2,\cdots,d^2\}$. Thus, the eigenvalue $\Gamma$ is $(d^2-1)$-fold degenerate, and the eigenvalue $\Lambda_1$ is non-degenerate. The real-valued $NM\times d^2$ matrix 
$X_{i,\mu}$  consists of $d^2$ $NM$-dimensional orthonormal  arrays, i.e.
\begin{eqnarray}
\sum_{i=1}^{NM} X_{i,\mu} X_{i,\nu} &=& \delta_{\mu \nu}
\label{bas}
\end{eqnarray}
for $\mu,\nu \in \{1,\cdots,d^2\}$. They fulfill the relations
\begin{eqnarray}
X_{i,1} &=&\frac{1}{\sqrt{NM}},~~
\sum_{a=1}^M X_{i(\alpha,a),\mu}
= 0 
\label{sum1}
\end{eqnarray}
for $\mu \in \{2,\cdots,d^2\}$.
As a consequence of the defining constraints (\ref{add1})  and (\ref{add2})
of $(N,M)$-POVMs
these orthonormal $NM$-dimensional  arrays also fulfill the relation
\begin{eqnarray}
\sum_{\mu=2}^{d^2} (X_{i,\mu})^2 &=&\frac{M-1}{M}.
\label{ineq1}
\end{eqnarray}

It is apparent from (\ref{S2})  that all basis operators ${G}_{\mu}$ with $\mu \in \{2,\cdots,d^2\}$ are mapped conformally onto a $(d^2-1)$-dimensional subspace of ${\cal H}_{NM}$ by stretching the norms of all its elements by the factor $\sqrt{\Gamma}$. Only the basis operator ${G}_1$ is stretched by a different factor, namely $\sqrt{\Lambda_1}$. 
Therefore, ignoring the positive semidefiniteness constraints the defining properties of $(N,M)$-POVMs (\ref{add2}) and (\ref{add3}) imply
the basis expansion
\begin{eqnarray}
\Pi_{i(\alpha,a)} &=& \frac{\unit_d}{M} + \sqrt{\Gamma} \sum_{\mu=2}^{d^2}  X_{i(\alpha,a),\mu} G_{\mu}
\label{basisexpansion2}
\end{eqnarray}
for each element of an informationally complete $(N,M)$-POVM
 with $i\in \{1,\cdots,NM\}$.
In view of
 (\ref{operatorbasis})
 the orthonormal hermitian operators $G_{\mu}$ for $\mu \in \{2,\cdots,d^2\}$
 are only determined up to an orthogonal transformation of the orthogonal group $O(d^2-1)$. Furthermore, there is an additional freedom in choosing the $N\times d^2$ matrices $X_{i(\alpha,a)}$ within the 
constraints imposed by relation (\ref{sum1}).
From the basis expansion (\ref{basisexpansion2}) it is apparent that  possibilities for constructing positive semidefinite POVM elements  may be severely constrained by not fully taking advantage of the freedom of choice of the orthonormal hermitian operator basis.

According to   relation (\ref{dimension5}) an informationally complete $(N,M)$-POVM consists of $d^2 -1 =  N(M-1)$ linear independent POVM elements.
A strategy  for its construction
 is to partition the orthonormal traceless hermitian operator basis $\{G_{\mu}\}$ with $\mu \in \{2,\cdots, d^2\}$ into $N$ basis tuples $B_{\alpha}$, each of which corresponds to a particular value of $\alpha\in \{1,\cdots,N\}$. This partitioning of the basis elements ensures that condition (\ref{add3}) is fulfilled. Accordingly, the basis expansion (\ref{basisexpansion2}) is restricted to an ansatz of the form
\begin{eqnarray}
\Pi_{i(\alpha,a)} &=& \frac{\unit_d}{M} + \sqrt{\Gamma} \sum_{G_{\mu}\in B_{\alpha}} X_{i(\alpha,a),\mu} G_{\mu}.
\label{basisexpansion2a}
\end{eqnarray}
Using this ansatz the allowed transformations are restricted to the orthogonal group $O(M-1)$ for each value of $\alpha$, thus also restricting the achievable positive semidefinite operators for a given basis $\{G_\mu\}$. Therefore, the possible values of the continuous parameters $x$ of such a construction  depend on the chosen basis $\{G_\mu\}$ and its partitioning.
In order to demonstrate this, let us consider the construction of a MUM for $d=3$ as a special example of an informationally complete $(4,3)$-POVM with $1/3< x\leq 1$.
According to (\ref{basisexpansion2a})  the  construction of this MUM can be interpreted geometrically. For this purpose let us identify the operator $\unit_d/3$ with the origin of an $8$-dimensional Euclidean space spanned by the hermitian operators $G_{\mu}$ for $\mu\in\{2,\dots,9\}$. Accordingly, we have to construct $N=4$ equilateral triangles with this origin as their centroids to fulfill the characteristic completeness relation of POVMs (\ref{sum1}) for each $\alpha$. Furthermore, condition (\ref{maximal}) implies that
\begin{align}
	\Gamma \sum_ {G_{\mu} \in B_{\alpha}}  (X_{i(\alpha,a),\mu})^2\leq r_>^2:= 2/3\label{rnm}.
\end{align}
A $(N,M)$-POVM is optimal if the two sides of inequality (\ref{rnm}) are equal.

\begin{figure}
\includegraphics[width=0.32\linewidth,
height=0.23\textheight]{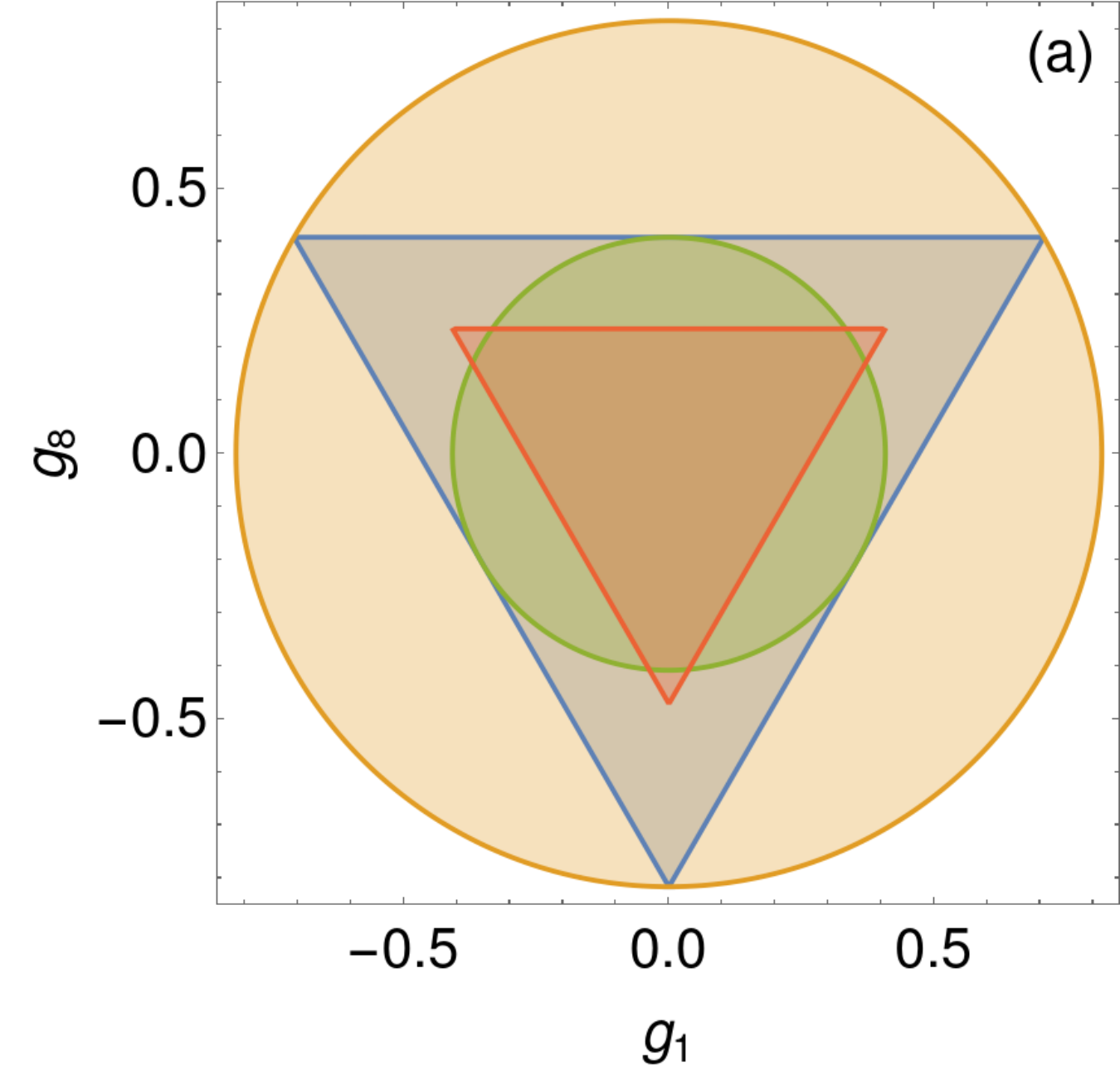}
\includegraphics[width=0.32\linewidth,
height=0.23\textheight]{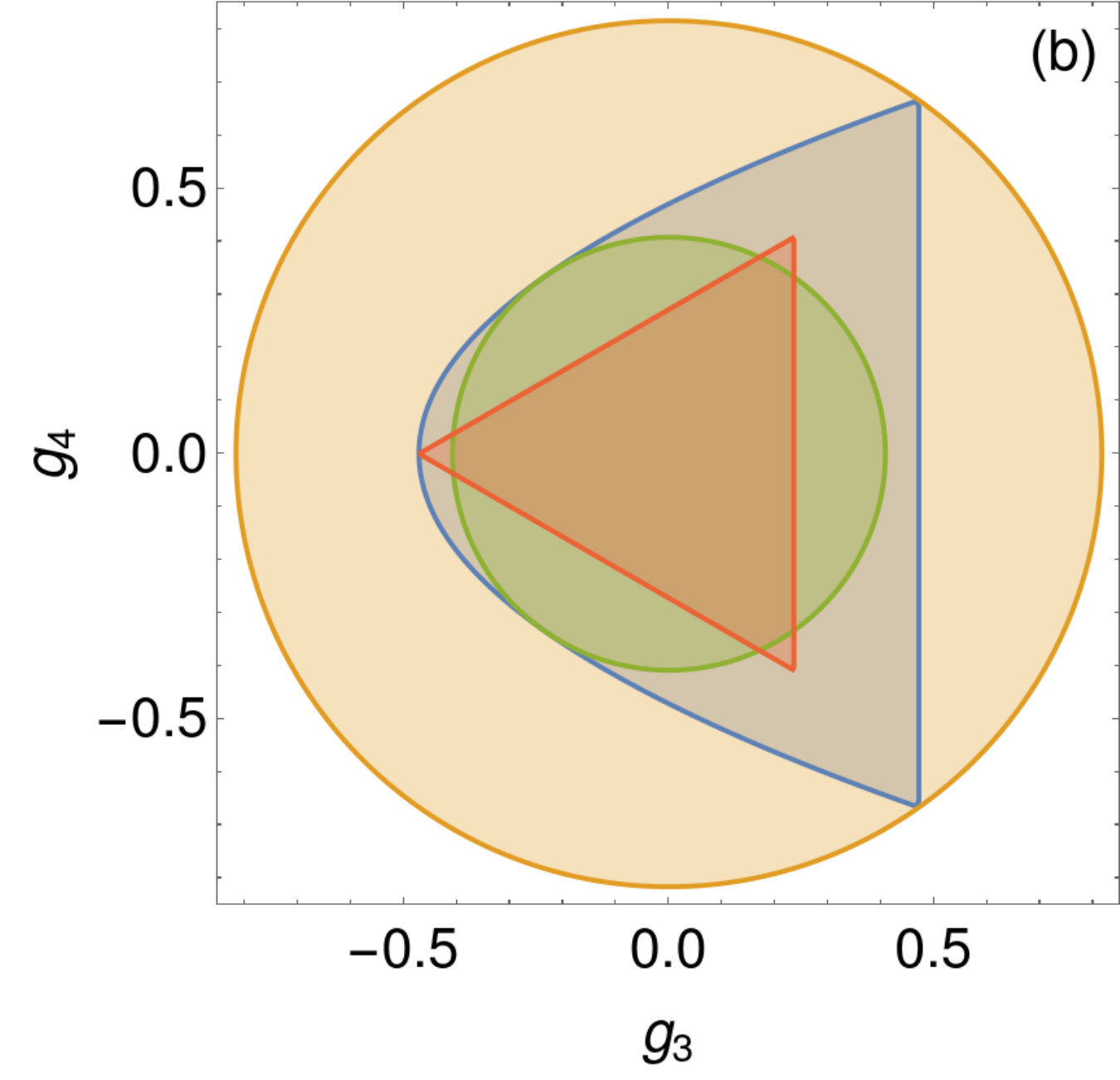}
\includegraphics[width=0.32\linewidth,
height=0.23\textheight]{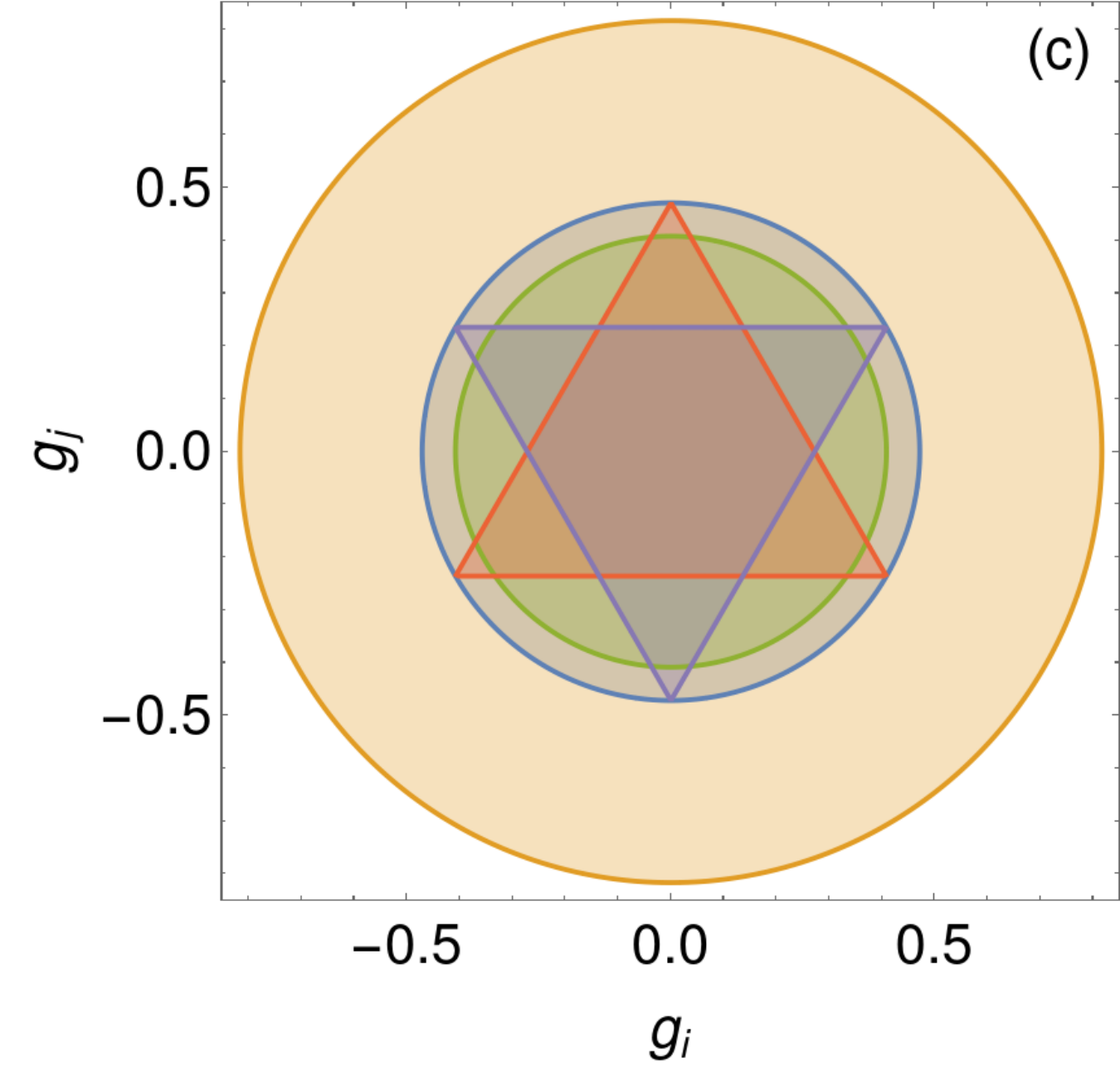}
\caption{Regions of positive semidefiniteness corresponding to four different partitions $B_{\alpha}$, $\alpha\in \{1,\cdots,4\}$ of the traceless, hermitian operators for an informationally complete $(4,3)$-POVM (MUM) in dimension $d=3$ with $\alpha=1$ (a), $\alpha=2$ (b), $\alpha=\{3,4\}$ and $(i,j) \in \{(2,5),(6,7)\}$ (c):  Geometrically each equilateral triangle with $(0,0)$ as its centroid represents a set of three operators fulfilling (\ref{basisexpansion2a}). The restrictions imposed by positive semidefiniteness are visualized by the blue regions. The yellow regions represent the constraints  (\ref{maximal}). The green region is the circle of maximal radius $r_<=1/\sqrt{6}$ with center $(0,0)$ located within the intersection of all blue regions of all four partitions. Within this circle equilateral triangles with centroid $(0,0)$ can be rotated by arbitrary angles. The two red equilateral triangles of Fig.\ref{Fig.1-3}a-b and the two equilateral triangles of 
Fig.\ref{Fig.1-3}c
represent a possible maximal MUM with $x=5/9$.  Their vertices lie outside of the green circles. 
}
\label{Fig.1-3}
\end{figure}

In Fig.\ref{Fig.1-3} the constraints imposed by the positive semidefiniteness  of all POVM elements of this MUM are visualized graphically for the different partitions $B_{\alpha}$. Thereby, the Gell-Mann matrices as defined in appendix A have been used as an orthonormal basis of traceless, hermitian operators. Accordingly, the four partitions have been chosen as
$B_1 = \{g_1,g_8\}, B_2 = \{g_3,g_4\},B_3 = \{g_2,g_5\},B_4 = \{g_6,g_7\}$. For arbitrary partitionings the corresponding positive semidefinite regions have already been discussed recently \cite{Kimura}. In
Fig.\ref{Fig.1-3}a the two-dimensional Euclidean subspace spanned by the unit vectors of partition $B_1$ is depicted. All points  inside the blue triangle correspond to the convex set of positive semidefinite matrices according to (\ref{basisexpansion2a}). The vertices of any equilateral triangle within this blue triangle with the origin as its centroid constitute a triple of possible POVMs with $a\in \{1,\cdots,3\}$ and $\alpha=1$. The points inside the yellow circle correspond to all hermitian operators, which fulfill the necessary constraint (\ref{maximal}). 
From inequality (\ref{rnm}) it follows that an optimal POVM is an equilateral triangle, whose vertices are on the boundary of the yellow area so that they have maximal distance to the origin.  The blue triangle itself constitutes a single optimal POVM, which can be constructed with the help of the partitioning $B_1$ of the Gell-Mann basis.
The green circle is the maximal circle around the origin, which can be constructed inside the triangle of positive semidefinite elements. Its radius is given by $r_< = 1/\sqrt{6}$. The blue region of Fig.\ref{Fig.1-3}b shows the convex set of positive semidefinite hermitian matrices, which can be constructed in
the two-dimensional Euclidean subspace spanned by unit vectors of partition $B_2$.  The vertices of any equilateral triangle constructed within this blue region with the origin as its centroid constitute a triple of possible POVM elements with $a\in \{1,\cdots,3\}$ and $\alpha=2$. Again, the points inside the yellow circle correspond to all hermitian matrices, which fulfill the necessary constraint (\ref{maximal}). The two points of the blue area intersecting with the yellow area's boundary cannot be used for constructing an optimal POVM for $\alpha=2$. The green circle is the maximal circle around the origin again with radius $r_< = 1/\sqrt{6}$, which can be constructed inside the blue region of positive semidefinite elements. 
The blue region of Fig.\ref{Fig.1-3}c shows the convex set of positive, semidefinite, hermitian matrices, which can be constructed in
the two-dimensional Euclidean subspace spanned by unit vectors of partitions $B_3$ or $B_4$.  The vertices of any equilateral triangle  constructed within this blue region with the origin as its centroid constitute a triple of possible POVMs with $a\in \{1,\cdots,3\}$ and $\alpha=3$ or $\alpha=4$. Again the points inside the yellow circle correspond to all hermitian matrices fulfilling the necessary constraint (\ref{maximal}). A MUM is given by four equilateral triangles of identical size inside the positive semidefinite area of each partition with the origin as its centroid. The red triangles in Figs.\ref{Fig.1-3}a-b
and the two triangles in Fig.\ref{Fig.1-3}c show four equilateral triangles  of maximal sizes, which can be constructed inside the blue regions of Figs.\ref{Fig.1-3}a-c. These four triangles constitute an informationally complete MUM in dimension $d=3$ with the maximal possible value of $x=5/9$.  The directions of some of these triangles  with respect to the chosen partitioning of the Gell-Mann basis are not determined uniquely. In Fig.\ref{Fig.1-3}a, for example the red triangle can be rotated around its centroid continuously as long as it stays within the blue region of positive semidefiniteness. In particular, this implies that not all rotation angles are possible. However, in Fig.\ref{Fig.1-3}c rotations of the two triangles around the origin are possible for arbitrary angles. But in Fig.\ref{Fig.1-3}b the shape of the blue region  implies that the position of the red triangle is fixed uniquely. From Figs.\ref{Fig.1-3}a-c it is apparent that 
the vertices of the equilateral triangles representing the maximal informationally complete  MUM are located outside of the green circle of radius $r_< = 1/\sqrt{6}$. 
In contrast, all POVM elements, whose equilateral triangles are constructed inside this green circle, can be rotated arbitrarily around the origin without affecting the positive semidefiniteness of their corresponding POVM elements.

In general
it is a complicated task to determine criteria, under which all POVM elements of a $(N,M)$-POVM are positive semidefinite. This example demonstrates that, although for sufficiently small regions  around the minimal possible value of $x=d/M^2$ $(N,M)$-POVMs can be constructed, complications increase with increasing values of $x$. Typically, the most complicated situations arise for constructions of optimal $(N,M)$-POVMs. In view of these difficulties it is of interest to develop conditions for the existence of optimal $(N,M)$-POVMs. Motivated by this need in the following such conditions will be developed.

\section{A sufficient condition for the construction of $(N,M)$-POVMs\label{SecIII}}
In this section a sufficient condition is derived, under which for a $d$-dimensional quantum system $(N,M)$-POVMs  can always be constructed. This sufficient condition yields a simple upper bound on the continuous $x$-parameter (cf. inequality (\ref{ineq2})), within which this can be achieved.

\begin{figure}
\includegraphics[width=0.7\linewidth,height=0.42\textheight]{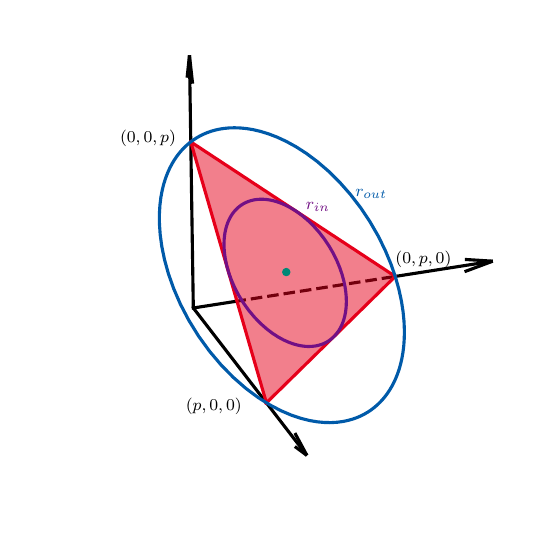}
\caption{Visualization of the simplices $\Delta_{d-1}$ and $\Delta_{d-2}=\partial \Delta_{d-1}$ and of their corresponding centroids  in the elementary case $d=3$: The orthogonal axes $x,y,z$ are defined by the one-dimensional orthogonal projection operators $\{P_{\sigma}\mid \sigma \in \{1,2,3\}\}$ and $p= \Tr\{\Pi_{i(\alpha,a)}\}=d/M$. The simplex $\Delta_2$ is a triangle (red area). Its centroid $C_2$ is represented by the green point.
The $1$-simplex $\Delta_1 = \partial \Delta_2$ is the boundary of the red triangle. 
The dark blue circle centered around $C_2$ with radius $r_{in}$ is the maximal circle, which can be constructed inside $\Delta_2$. It touches $\Delta_1$ in its three centroids. The light blue circle centered around $C_2$  with radius $r_{out}$ represents the constraint (\ref{maximal}).}
\label{simplex}
\end{figure}

In general it is a complicated task to determine criteria, under which all POVM elements of a $(N,M)$-POVM  of the form (\ref{basisexpansion2}) are positive semidefinite. However, a general sufficient condition for positivity can be derived by using general properties of positive semidefinite linear operators \cite{geometry-bengtsson, Kossakowski}. For this purpose let us consider an arbitrary POVM element of a $(N,M)$-POVM in a $d$-dimensional Hilbert space. Its spectral representation is given by
\begin{eqnarray}
\Pi_{i(\alpha,a)} &=&\sum_{\sigma=1}^{d} \lambda_{\sigma} P_{\sigma}
\label{POVMelement}
\end{eqnarray}
with its non-negative eigenvalues $\lambda_{\sigma}$ 
and
with the associated one-dimensional orthogonal projection operators $P_{\sigma}$ fulfilling the completeness and orthogonality relations $\sum_{\sigma=1}^{d} P_{\sigma} = \unit_d$ and $P_{\sigma}P_{\sigma'} = \delta_{\sigma, \sigma'} P_{\sigma}$.
The constraint (\ref{add1}) yields the relation
\begin{eqnarray}
\Tr\{\Pi_{i(\alpha,a)}\} &=&\frac{d}{M}= \sum_{\sigma=1}^{d} \lambda_{\sigma}.
\label{Pi-element}
\end{eqnarray}
Therefore, for given projection operators $P_{\sigma}$
the set of all positive semidefinite POVM elements of this $(N,M)$-POVM constitute a $(d-1)$-dimensional simplex $\Delta_{d-1}$  in the $d$-dimensional Hilbert space (compare with Fig.\ref{simplex}). The centroid of this $(d-1)$-simplex is given by
\begin{eqnarray}
C_{d-1} &=& \frac{\Tr\{\Pi_{i(\alpha,a)}\}}{d}\sum_{\sigma=1}^d P_{\sigma} = 
\frac{1}{M}\unit_d.
\label{Cd1}
\end{eqnarray}
The boundary of this $(d-1)$-simplex is a $(d-2)$-simplex, i.e. $\Delta_{d-2}:= \partial \Delta_{d-1}$.
 The boundary of $\Delta_{d-1}$ consists of all possible elements of the form (\ref{POVMelement}) with
at least one of the $d$ eigenvalues vanishing. The centroid $C_{d-1}$ has equal distances $r_{in}$ to the centroids of all the $d$ parts of $\Delta_{d-2}$. This distance $r_{in}$ defines the radius of the largest possible circle with center $C_{d-1}$, which lies within $\Delta_{d-1}$ and touches $\Delta_{d-2}$ in one of its $d$ centroids $C_{d-2}$. It is determined by the relation
\begin{eqnarray}
r_{in}^2&=& \Tr\{(C_{d-1}-C_{d-2})^2\} = (\Tr\{\Pi_{i(\alpha,a)}\} )^2\left[(d-1)\left(\frac{1}{d}-\frac{1}{d-1} \right)^2 +\frac{1}{d^2}\right] = \frac{(\Tr\{\Pi_{i(\alpha,a)}\} )^2}{d(d-1)} = \frac{d}{M^2(d-1)}.
\label{r1}
\end{eqnarray}
Using (\ref{add1}) and (\ref{add2}) 
we arrive at the inequality 
\begin{eqnarray}
0 < \Tr\{(\Pi_{i(\alpha,a)} - \unit_d/M)^2\} = x-\frac{d}{M^2}\leq r_{in}^2= \frac{d}{M^2(d-1)}.
\label{ineq2}
\end{eqnarray}
According to (\ref{Cd1}) and (\ref{r1}) the centroid $C_{d-1}$ as well as the radius $r_{in}$ are independent of the choice of the orthonormal projection operators $P_{\sigma}$ so that (\ref{ineq2}) applies to all POVM elements.
Therefore,
it can be concluded that fulfillment of inequality (\ref{ineq2}) is sufficient for the existence of a $(N,M)$-POVM. It guarantees the positive semidefiniteness of all its POVM elements. Note that in the special cases considered in Figs.\ref{Fig.1-3}a-c, i.e. $M=3, d=3$, the distance
$r_{in}$ reduces to the value 
$r_{in} = 1/\sqrt{6}=r_{<}$, which is a basis and partition independent value.

One can also define a circle with center $C_{d-1}$ and radius $r_{out}$, within which all possible $(N,M)$-POVMs are located according to the constraint (\ref{maximal}). It is defined by
\begin{eqnarray}
r_{out}^2 &=& {\rm min}\left(\frac{d(M-1)}{M^2}, \frac{d(d-1)}{M^2} \right)
\end{eqnarray}
so that (\ref{maximal}) reduces to the relation $0< x-d/M^2 \leq r_{out}^2$.
A $(N,M)$-POVM with $x-d/M^2= r_{out}^2$ is called optimal. The ratio between the range of $x$-values around its minimal value of $d/M^2$, for which a $(N,M)$-POVM can always be constructed, i.e. $r_{in}^2$, and the corresponding maximal possible range, i.e. $r_ {out}^2$, is given by
\begin{eqnarray}
R(d) &=& \frac{r_{in}^2}{r_ {out}^2}  =
\left\{
\begin{array}{ll}
\frac{1}{(d-1)^2}& M \geq d\\
\frac{1}{(M-1)(d-1)}& 2 \leq M < d
\end{array}
\right. .
\label{R(d)}
\end{eqnarray}
\begin{figure}
\includegraphics[width=0.5\linewidth,
height=0.25\textheight]{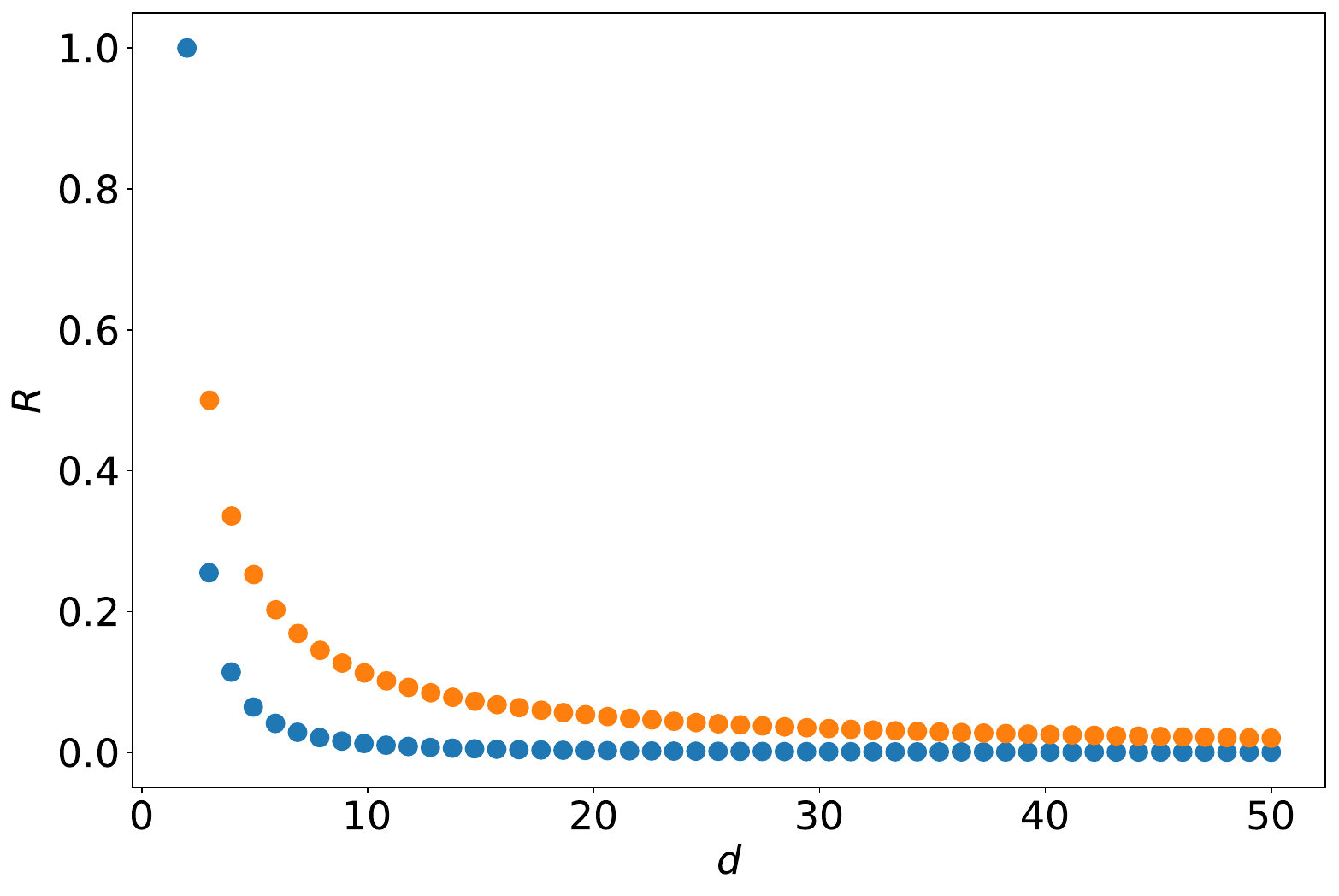}
\caption{Dimensional dependence of $R(d)$ according to (\ref{R(d)}) for $M\geq d$ (blue points) and for $2=M < d$ (orange points):  Cases with $2<M<d$ are located between these two series of points and also rapidly converge to zero with increasing dimensions $d$. 
}
\label{ratio}
\end{figure}

In Fig.\ref{ratio} the dependence of this ratio $R(d)$ is depicted for cases with $M\geq d$ (blue points) and for $2=M < d$ (orange points).
It is apparent that this ratio
converges to zero rapidly with increasing dimensions of the quantum system's Hilbert space $d$. Correspondingly, 
the size of the interval of $x$-values, for which  $(N,M)$-POVMs  can be constructed with arbitrary choices of traceless, orthonormal, hermitian operator bases, rapidly tends to zero. For cases with $2 < M < d$ the values of $R(d)$ are located inside the region between the two series of dotted points of Fig.\ref{ratio}.
In the exceptional case of a qubit, i.e. $d = 2$, 
the inner and outer radius are identical, i.e. $r_{in}^2=r_{out}^2=2/M^2$, and 
for $M = 4$ the set of positive semidefinite operators forms the
Bloch sphere.

Fulfillment of the sufficient condition (\ref{ineq2}) allows the construction of $(N,M)$-POVMs for arbitrary choices of the $(d^2-1)$ traceless elements of the hermitian operator basis of ${\cal H}_{d^2}$ according to the ansatz (\ref{basisexpansion2}). However, from the explicit expression of $r_{in}^2$ (cf. (\ref{r1}) and Fig.\ref{ratio})) it is also apparent that the range of $x$-values, for which this sufficient condition can be fulfilled, decreases rapidly with increasing values of $M$. Thus, in general it is an intricate problem to construct $(N,M)$-POVMs, if the sufficient condition of (\ref{ineq2}) is not applicable. In particular, in these cases the choice of the traceless hermitian operator basis elements entering (\ref{basisexpansion2}) can be crucial for the construction. Typically the most complicated situations arise for the  construction of  optimal $(N,M)$-POVMs. Motivated by these problems in the subsequent section we explore necessary conditions for the construction of optimal $(N,M)$-POVMs.

\section{Necessary conditions for the existence of  optimal
$(N,M)$-POVMs\label{Sec-necc}}
In this section two necessary conditions for the existence of optimal  $(N,M)$-POVMs of a $d$-dimensional quantum system are presented. As a first main result it is shown that for $M\geq d$ the existence of $(d^2-1)$ isospectral, traceless, orthonormal, hermitian operators is necessary for the existence of an optimal informationally complete $(N,M)$-POVM. The form of their common spectrum is completely determined by the defining properties of the optimal informationally complete $(N,M)$-POVM. As a second main result it is demonstrated that in cases with $2 < M <d$ the existence of an optimal
  $(N,M)$-POVM is only possible, if all its POVM elements are projection operators of equal rank.

\subsection{Optimal informationally complete $(N,M)$-POVMs for  $M\geq d$}

Let us consider an optimal informationally complete $(N,M)$-POVM $\Pi$ 
of a $d$-dimensional quantum system with $M\geq d$. 
 According to the defining relations (\ref{add1}) and (\ref{add2}), each element $\Pi_{i(\alpha,a)}$ 
of this $(N,M)$-POVM
fulfills the relations
\begin{eqnarray}
\Tr\{\Pi_{i(\alpha,a)}\} &=& \sum_{\sigma=1}^d \lambda_{\sigma} = \frac{d}{M},~~
\Tr\{(\Pi_{i(\alpha,a)})^2\} = \sum_{\sigma=1}^d \lambda_{\sigma}^2 = \frac{d^2}{M^2}
\end{eqnarray}
with non-negative eigenvalues $\lambda_{\sigma}$ for $\sigma\in \{1,\cdots,d\}$. Both relations constrain the possible values of these eigenvalues because they imply
$\sum_{1=\sigma < \sigma^{'}}^d \lambda_{\sigma} \lambda_{\sigma^{'}} = 0$. 
Therefore, the positive semidefiniteness of all eigenvalues $\lambda_{\sigma}$ for $\sigma\in \{1,\cdots,d\}$ implies that there is only one non-zero eigenvalue of magnitude $d/M$. Correspondingly, an arbitrary  POVM element of the optimal informationally complete $(N,M)$-POVM $\Pi$ has to be of the general form
\begin{eqnarray}
\Pi_{i(\alpha,a)} &=& \frac{d}{M} | i(\alpha,a) \rangle \langle i(\alpha,a) |.
\label{Pi-def}
\end{eqnarray}
The defining relations (\ref{add1}), (\ref{add2}) and (\ref{add3}) constrain the scalar products of the generally non-orthogonal but normalized eigenstates $|i(\alpha,a)\rangle$ by the relations
\begin{eqnarray}
\mid \langle  i(\alpha,a) | i(\alpha,a') \rangle \mid &=& \sqrt{\frac{M/d -1}{M-1}},~~
\mid \langle  i(\alpha,a) | i(\beta,b)\rangle \mid~ = ~\sqrt{\frac{1}{d}}
\end{eqnarray}
for $\alpha\neq \beta \in \{1,\cdots, N\}$, $a\neq a'\in \{1,\cdots, M\}$ and $a,a',b\in \{1,\cdots, M\}$. 
For each $\alpha \in \{1,\cdots, N\}$ and $a\in \{1,\cdots, M-1\}$
one can construct the traceless, hermitian operators
\begin{eqnarray}
G_{i(\alpha,a)} &=& \frac{M-1}{(\sqrt{M} +1) \sqrt{d^2 -d}}\left(\unit_d + A_{i(\alpha,a)}\right)~~{\rm with}~~
A_{i(\alpha,a)} =
\sqrt{M} \Pi_{i(\alpha,M)} -\sqrt{M}(\sqrt{M} +1) \Pi_{i(\alpha,a)}.
\label{ON-basis}
\end{eqnarray}
It is straightforward to demonstrate that these $(M-1)N = d^2-1$ (cf. (\ref{dimension5})) hermitian operators  $G_{i(\alpha,a)}$ are an orthonormal basis in the Hilbert space ${\cal H}_{d^2 -1}$ of traceless, hermitian operators of the $d$-dimensional quantum system. According to (\ref{Pi-def}) and (\ref{ON-basis}) each hermitian operator $A_{i(\alpha,a)}$ has a maximal rank of two. Therefore, the characteristic polynomials determining the eigenvalues $\Lambda$ of $A_{i(\alpha,a)}$ have the general form
\begin{eqnarray}
\Lambda^d + c_{d-1} \Lambda^{d-1} + c_{d-2} \Lambda^{d-2} &=&0.
\label{quadratic}
\end{eqnarray}
Using the Cayley-Hamilton theorem \cite{Cayley-Hamilton} it is easily found that
\begin{eqnarray}
c_{d-1} &=&-\tr\{A_{i(\alpha,a)}\} = d,~~
c_{d-2} ~=~\frac{1}{2}\left((\tr\{A_{i(\alpha,a)}\})^2-
\tr\{A_{i(\alpha,a)}^2
\} \right) = \frac{d - d^2}{\sqrt{M} -1}.
\end{eqnarray}
Consequently, all traceless, orthonormal, hermitian operators $G_{i(\alpha,a)}$ have the same spectrum ${\rm Sp}(G_{i(\alpha,a)})$
determined by the solutions of (\ref{quadratic}), i.e.
\begin{eqnarray}
{\rm Sp}(G_{i(\alpha,a)}) &=&\left\{\left(
\frac{\sqrt{M-1}}{(\sqrt{M}+1)\sqrt{d^2-d}}(1+\Lambda_+)\right)^{(1)},
\left(\frac{\sqrt{M-1}}{(\sqrt{M}+1)\sqrt{d^2-d}}(1+\Lambda_-)\right)^{(1)},
\left(\frac{\sqrt{M-1}}{(\sqrt{M}+1)\sqrt{d^2-d}}\right)^{(d-2)}
\right\}\nonumber\\
\label{spectrum1}
\end{eqnarray}
with 
\begin{eqnarray}
\Lambda_{\pm} &=&\frac{1}{2}\left( -d \pm \sqrt{d^2+4\frac{d^2-d}{\sqrt{M}-1}}\right).
\end{eqnarray}
The numbers in brackets in the exponents of (\ref{spectrum1}) indicate the multiplicities of the corresponding eigenvalues. 

Therefore, it can be concluded that for $M\geq d$ the existence of an optimal informationally complete $(N,M)$-POVM $\Pi$ implies the existence of a set of $(d^2-1)$ isospectral, traceless,  orthonormal
hermitian operators $G_{i(\alpha,a)}$ defined by (\ref{ON-basis}), whose common spectrum is given by (\ref{spectrum1}). Stated differently, the existence of a 
set of $(d^2-1)$ isospectral, traceless, orthonormal, hermitian operators $G_{i(\alpha,a)}$, whose common spectrum is given by (\ref{spectrum1}), is necessary for the existence of an optimal informationally complete $(N,M)$-POVM.
This result  generalizes an already known property of GSICs \cite{GSIC-POVM}, i.e. optimal informationally complete $(1,M)$-POVMs, to all optimal informationally complete $(N,M)$-POVMs with $M\geq d$.

\subsection{Optimal $(N,M)$-POVMs for
 $2<M<d$\label{Sec-2}}
In this subsection we prove a necessary condition for the existence of an optimal $(N,M)$-POVM for cases with $2 <M<d$. It is shown that in these cases it is necessary that all POVM elements are projection operators of rank $d/M$. This significantly constrains the possible parameters $(d,N,M)$, for which optimal $(N,M)$-POVMs can be constructed. Although the discussion of this necessary condition also applies to $M=2$, these special cases will be discussed separately in Sec.\ref{Sec-3}.

Let us consider an optimal $(N,M)$-POVM of a $d$-dimensional quantum system with $2 <M< d$.  According to (\ref{maximal}) it is characterized by the parameter $x=d/M$. Furthermore, the defining conditions (\ref{add1}) and (\ref{add2}) imply that each element $\Pi_{i(\alpha,a)}$ (cf. (\ref{POVMelement})) 
of this optimal $(N,M)$-POVM  $\Pi$ 
fulfills the relations
\begin{eqnarray}
\Tr\{\Pi_{i(\alpha,a)}\} &=& \sum_{\sigma=1}^d \lambda_{\sigma} = \frac{d}{M},~~
\Tr\{(\Pi_{i(\alpha,a)})^2\} ~=~ \sum_{\sigma=1}^d \lambda_{\sigma}^2 = \frac{d}{M}
\label{necc}
\end{eqnarray}
with non-negative eigenvalues $\lambda_{\sigma}$ for $\sigma\in \{1,\cdots,d\}$. Consequently, the only possible eigenvalues are given by
$\lambda_{\sigma} \in \{0,1\}$. Therefore, an optimal $(N,M)$-POVM can exist only in dimensions $d$, for which $d/M \in {\mathbb N}$ is a natural number. Furthermore, all POVM elements are projection operators of rank $d/M$, i.e. $\Pi^2_{i(\alpha,a)} = \Pi_{i(\alpha,a)}$. 
Stated differently, the existence of POVM elements of rank $d/M \in {\mathbb N}$, which are projection operators,
is necessary for the existence of an optimal $(N,M)$-POVM for $2< M <d$.
According to this necessary condition the smallest dimension $d$, for example, in which an optimal informationally complete $(N,M)$-POVM can possibly be constructed, is given by $d=8$. It is a $(21,4)$-POVM with $x=2$ and all its POVM elements are of rank two.

\section{Optimal $(N,2)$-POVMs\label{Sec-3}}
In this section optimal  $(N,2)$-POVMs of $d$-dimensional quantum systems are investigated. For them additional properties can be derived, which transcend the necessary condition discussed in Sec.\ref{Sec-2}. These additional properties are strong enough for deriving a necessary and sufficient condition for the existence of optimal $(N,2)$-POVMs with $N\leq d^2-1$, including informationally complete ones for $N=d^2-1$. 

Let us consider an arbitrary POVM element of an optimal $(N,2)$-POVM  as given by (\ref{POVMelement})  with $x=d/2$ and positive semidefinite eigenvalues of the form $\lambda_{\sigma}=1/2 +\eta_{\sigma}$ for 
$\sigma\in \{1,\cdots, d\}$. 
According to the arguments of Sec.\ref{Sec-2} $\lambda_{\sigma} \in \{0,1\}$ so that $\mid \eta_{\sigma} \mid~ = ~1/2$. In addition, we have $d/2 \in {\mathbb N}$ so that the dimension $d$  has to be even. Therefore, relation (\ref{necc}) can only be fulfilled, if the spectrum of the traceless, normalized and hermitian operators 
\begin{eqnarray}
K_{i(\alpha,a)} &=& \frac{ \Pi_{i(\alpha,a)} - \unit_d/2 }{\sqrt{d/4}}
\label{Ki}
\end{eqnarray}
 is given by
\begin{eqnarray}
{\rm Sp}\left(K_{i(\alpha,a)}\right) &=&
\left\{+\frac{1}{\sqrt{d}}^{(d/2)}, - \frac{1}{\sqrt{d}}^{(d/2)} \right\}
\label{spectrum2}
\end{eqnarray}
for each $i(\alpha,a) \in \{1,\cdots, 2N\}$. 
The numbers in brackets in the exponents of (\ref{spectrum2}) indicate the multiplicities of the corresponding eigenvalues. 
In view of relation (\ref{add3})
the operators $K_{i(\alpha,a)}$ and $K_{i(\beta,b)}$ are also orthogonal for $\alpha \neq \beta \in \{1,\cdots,N\leq d^2-1 \}$ and $a,b\in \{1,2\}$. Thereby, we have taken into account that for a  $d$-dimensional quantum system the number of  traceless, orthogonal, hermitian operators cannot exceed $d^2-1$. But these operators $K_{i(\alpha,a)}$ and $K_{i(\beta,b)}$ are not orthogonal for $\alpha=\beta$ and $a\neq b$ and fulfill the relation
\begin{eqnarray}
K_{i(\alpha,2)}&=&-K_{i(\alpha,1)}.
\label{-Ki}
\end{eqnarray}
Therefore, for $N\leq d^2- 1$ the existence of an optimal $(N,2)$-POVM implies the existence of $N$ isospectral, traceless, orthonormal, hermitian operators $\{K_{i(\alpha,1)}\mid \alpha \in \{1,\cdots,N\}\}$, whose common spectrum is given by (\ref{spectrum2}). However, in view of (\ref{Ki}) and (\ref{-Ki}) this conclusion can also be turned around.
Thus, for $N\leq d^2- 1$ the existence of an optimal $(N,2)$-POVM is sufficient and necessary for the existence of $N$ isospectral, traceless, orthonormal, hermitian operators $\{K_{i(\alpha,1)}\mid \alpha \in \{1,\cdots,N\}\}$, whose common spectrum is given by (\ref{spectrum2}). Thereby, the case $N=d^2-1$ covers optimal informationally complete $(N,2)$-POVMs.
It should be mentioned that this existence criterion for optimal $(N,2)$-POVMs with $N\leq d^2-1$  generalizes a recent weaker result
 \cite{NMPOVM}, which was based  on the weaker assumption $\mid \eta_{\sigma}\mid \leq 1/2$.

In the special cases of even dimensions of the form $d= 2^k$ with $k \in {\mathbb N}$,  $N\leq d^2-1$ isospectral, traceless, orthonormal, hermitian operators can easily be constructed with the help of
the Clifford algebra generated by tensor products of Pauli operators. Accordingly, these operators are given by 
\begin{eqnarray}
\frac{1}{\sqrt{2^k}}\sigma_{i_1}\otimes \sigma_{i_2}\otimes \cdots \sigma_{i_k}
\end{eqnarray}
with $(i_1,\cdots,i_k) \neq (0,\cdots,0)$ and
with the Pauli operators 
\begin{eqnarray}
\sigma_0 &=&\left(
\begin{array}{cc}
1&0\\
0&1
\end{array}
\right),~~
\sigma_1 =\left(
\begin{array}{cc}
0&1\\
1&0
\end{array}
\right),~~
\sigma_2 = \left(
\begin{array}{cc}
0&-i\\
i&0
\end{array}
\right),~~
\sigma_3 =\left(
\begin{array}{cc}
1&0\\
0&-1\end{array}
\right).
\end{eqnarray}
Therefore, for $N\leq d^2-1$ optimal $(N,2)$-POVMs in dimension $d=2^k$ can easily be constructed. Their $2N$ elements are given by
\begin{eqnarray}
\Pi_{i((i_1,\cdots,i_k),1)} &=&\frac{\unit_d}{2} + \frac{1}{2}
\sigma_{i_1}\otimes \sigma_{i_2}\otimes \cdots \sigma_{i_k},~~
\Pi_{i((i_1,\cdots,i_k),2)} ~=~\frac{\unit_d}{2} - \frac{1}{2}
\sigma_{i_1}\otimes \sigma_{i_2}\otimes \cdots \sigma_{i_k}
\end{eqnarray}
with 
$(i_1,\cdots,i_k) \neq (0,\cdots,0)$.
Nevertheless, this explicit basis construction still leaves the question open, to which extent such optimal $(N,2)$-POVMs also exist in other even dimensions.
 \section{Summary and conclusions}
 Motivated by the recent interest in basic theoretical properties of $(N,M)$-POVMs, we have explored 
general features of their existence and construction with the help of orthonormal, hermitian operator bases for arbitrary $d$-dimensional quantum systems.
A sufficient condition has been derived for the existence of an arbitrary $(N,M)$-POVM. It generalizes the already known property, that $(N,M)$-POVMs can always be constructed for sufficiently small values of their continuous $x$-parameter \cite{NMPOVM}. In particular, it yields an explicit expression for an upper bound on this continuous $x$-parameter, below which all POVM elements are guaranteed to be positive semidefinite. Also necessary conditions for the existence of  optimal $(N,M)$-POVMs have been presented. One of these necessary conditions exhibits a close connection between the existence of optimal informationally complete $(N,M)$-POVMs and the existence of isospectral,  traceless, orthonormal, hermitian operator bases in cases with $M\geq d$.
Thereby, we have built on recent results, which have established that the relation between $(N,M)$-POVMs and orthonormal hermitian operator bases is necessarily governed by highly degenerate linear maps \cite{Schumacher-Alber-PRA}. This necessary condition generalizes a property, recently found for the special case of GSICs \cite{GSIC-POVM}, to arbitrary optimal informationally complete $(N,M)$-POVMs.
This connection motivates further research on the construction of such isospectral, traceless, orthonormal hermitian operator  bases in order to shed new light on the construction of optimal informationally complete $(N,M)$-POVMs. 
Another necessary condition has been derived for optimal $(N,M)$-POVMs for cases with $M < d$. In particular, it has been shown that in these cases
all POVM elements necessarily have to be projection operators of equal rank. This significantly constrains the possible parameters for constructing optimal $(N,M)$-POVMs.
This necessary condition has been derived solely by using properties, which have to be fulfilled necessarily by all POVM elements. Therefore, it motivates further research on  conditions, which also take into account additional relations between  the different POVM elements of an optimal $(N,M)$-POVM.
For the special cases with $M=2$ a necessary and sufficient condition has been derived for the existence of optimal $(N,2)$-POVMs with $N\leq d^2-1$. Thereby, a relation to the existence  of a set of $N$ isospectral, traceless, orthonormal, hermitian operators has been established. Such operators with the required spectrum can only exist in even dimensions.
For dimensions $d=2^k,~k\in {\mathbb N}$ these operators can easily be constructed with the help of the Clifford algebra generated by the $k$-fold tensor products of the Pauli operators . 

The recently introduced $(N,M)$-POVMs \cite{NMPOVM} are potentially interesting for numerous tasks of quantum information processing, such as the exploration of provable entanglement in quantum communication or quantum state tomography. Our presented sufficient and necessary conditions do not only shed new light on currently open questions concerning their existence and construction but also concerning their application for practical purposes.
Our presented sufficient condition for their existence, for example, has established an explicit upper bound on the continuous $x$-parameters guaranteeing their existence. Combining this result with the recent observation \cite{Schumacher-Alber-PRA}, that typical bipartite entanglement can be detected locally in an optimal way by local $(N,M)$-POVMs fulfilling this sufficient condition, suggests interesting applications of $(N,M)$-POVMs for the detection of provable entanglement in quantum communication protocols. In view of these promising aspects also for applications we are confident that $(N,M)$-POVMs will play an interesting and practically useful role in future work exploring the intricacies of quantum correlations. 

\begin{acknowledgments}
G.A. is grateful to
his friend and regular collaborator A.R.P.Rau for numerous inspiring discussions on symmetries in quantum physics and beyond. 
It is a pleasure to dedicate this work to him. This research is supported by the Deutsche Forschungsgemeinschaft (DFG) -- SFB 1119 -- 236615297.
\end{acknowledgments}

\appendix
\section{A Gell-Mann basis for $d=3$}
The Gell-Mann basis, which has been used in obtaining Fig.\ref{Fig.1-3}a-c, is defined by the matrices
\begin{eqnarray}
g_1 &=&\frac{1}{\sqrt{2}}\left(
\begin{array}{ccc}
0&1&0\\
1&0&0\\
0&0&0
\end{array}
\right),~~
g_2 = \frac{1}{\sqrt{2}}\left(
\begin{array}{ccc}
0&-i&0\\
i&0&0\\
0&0&0
\end{array}
\right),~~
g_3 = \frac{1}{\sqrt{2}}\left(
\begin{array}{ccc}
1&0&0\\
0&-1&0\\
0&0&0
\end{array}
\right),~~
g_4 = \frac{1}{\sqrt{2}}\left(
\begin{array}{ccc}
0&0&1\\
0&0&0\\
1&0&0
\end{array}
\right),\nonumber\\
g_5 &=&\frac{1}{\sqrt{2}}\left(
\begin{array}{ccc}
0&0&-i\\
0&0&0\\
i&0&0
\end{array}
\right),~~
g_6 = \frac{1}{\sqrt{2}}\left(
\begin{array}{ccc}
0&0&0\\
0&0&1\\
0&1&0
\end{array}
\right),~~
g_7 = \frac{1}{\sqrt{2}}\left(
\begin{array}{ccc}
0&0&0\\
0&0&-i\\
0&i&0
\end{array}
\right),~~
g_8 = \frac{1}{\sqrt{6}}\left(
\begin{array}{ccc}
1&0&0\\
0&1&0\\
0&0&-2
\end{array}
\right).
\end{eqnarray}
These $(d^2-1)=8$ matrices have vanishing traces and are orthogonal with respect to the Hilbert-Schmidt scalar product. Together with the properly normalized unit matrix they form an orthonormal basis of the Hilbert space ${\cal H}_{d^2}$ for $d=3$.
\bibliographystyle{apsrmp4-1}
\bibliography{CJPRau}

\begin{thebibliography}{16}%
\makeatletter
\providecommand \@ifxundefined [1]{%
 \@ifx{#1\undefined}
}%
\providecommand \@ifnum [1]{%
 \ifnum #1\expandafter \@firstoftwo
 \else \expandafter \@secondoftwo
 \fi
}%
\providecommand \@ifx [1]{%
 \ifx #1\expandafter \@firstoftwo
 \else \expandafter \@secondoftwo
 \fi
}%
\providecommand \natexlab [1]{#1}%
\providecommand \enquote  [1]{``#1''}%
\providecommand \bibnamefont  [1]{#1}%
\providecommand \bibfnamefont [1]{#1}%
\providecommand \citenamefont [1]{#1}%
\providecommand \href@noop [0]{\@secondoftwo}%
\providecommand \href [0]{\begingroup \@sanitize@url \@href}%
\providecommand \@href[1]{\@@startlink{#1}\@@href}%
\providecommand \@@href[1]{\endgroup#1\@@endlink}%
\providecommand \@sanitize@url [0]{\catcode `\\12\catcode `\$12\catcode
  `\&12\catcode `\#12\catcode `\^12\catcode `\_12\catcode `\%12\relax}%
\providecommand \@@startlink[1]{}%
\providecommand \@@endlink[0]{}%
\providecommand \url  [0]{\begingroup\@sanitize@url \@url }%
\providecommand \@url [1]{\endgroup\@href {#1}{\urlprefix }}%
\providecommand \urlprefix  [0]{URL }%
\providecommand \Eprint [0]{\href }%
\providecommand \doibase [0]{http://dx.doi.org/}%
\providecommand \selectlanguage [0]{\@gobble}%
\providecommand \bibinfo  [0]{\@secondoftwo}%
\providecommand \bibfield  [0]{\@secondoftwo}%
\providecommand \translation [1]{[#1]}%
\providecommand \BibitemOpen [0]{}%
\providecommand \bibitemStop [0]{}%
\providecommand \bibitemNoStop [0]{.\EOS\space}%
\providecommand \EOS [0]{\spacefactor3000\relax}%
\providecommand \BibitemShut  [1]{\csname bibitem#1\endcsname}%
\let\auto@bib@innerbib\@empty
\bibitem [{\citenamefont {Bengtsson}\ and\ \citenamefont
  {Zyczkowski}(2006)}]{geometry-bengtsson}%
  \BibitemOpen
  \bibfield  {author} {\bibinfo {author} {\bibnamefont {Bengtsson},
  \bibfnamefont {I.}}, \ and\ \bibinfo {author} {\bibfnamefont
  {K.}~\bibnamefont {Zyczkowski}}} (\bibinfo {year} {2006}),\ \href@noop {}
  {\emph {\bibinfo {title} {Geometry of Quantum States}}}\ (\bibinfo
  {publisher} {Cambridge University Press},\ \bibinfo {address}
  {Cambridge})\BibitemShut {NoStop}%
\bibitem [{\citenamefont {Bergou}\ \emph {et~al.}(2021)\citenamefont {Bergou},
  \citenamefont {Hillery},\ and\ \citenamefont {Saffman}}]{POVM2}%
  \BibitemOpen
  \bibfield  {author} {\bibinfo {author} {\bibnamefont {Bergou}, \bibfnamefont
  {J.~A.}}, \bibinfo {author} {\bibfnamefont {M.~S.}\ \bibnamefont {Hillery}},
  \ and\ \bibinfo {author} {\bibfnamefont {M.}~\bibnamefont {Saffman}}}
  (\bibinfo {year} {2021}),\ \href@noop {} {\emph {\bibinfo {title} {Quantum
  Information Processing: Theory and Implementation}}}\ (\bibinfo  {publisher}
  {Springer},\ \bibinfo {address} {Cham})\BibitemShut {NoStop}%
\bibitem [{\citenamefont {Frobenius}(1878)}]{Cayley-Hamilton}%
  \BibitemOpen
  \bibfield  {author} {\bibinfo {author} {\bibnamefont {Frobenius},
  \bibfnamefont {G.}}} (\bibinfo {year} {1878}),\ \href@noop {} {\bibfield
  {journal} {\bibinfo  {journal} {J. Reine Angew. Math.}\ }\textbf {\bibinfo
  {volume} {84}},\ \bibinfo {pages} {1}}\BibitemShut {NoStop}%
\bibitem [{\citenamefont {Fuchs}\ \emph {et~al.}(2017)\citenamefont {Fuchs},
  \citenamefont {Hoang},\ and\ \citenamefont {Stacey}}]{GSIC-construction}%
  \BibitemOpen
  \bibfield  {author} {\bibinfo {author} {\bibnamefont {Fuchs}, \bibfnamefont
  {C.~A.}}, \bibinfo {author} {\bibfnamefont {M.~C.}\ \bibnamefont {Hoang}}, \
  and\ \bibinfo {author} {\bibfnamefont {B.~C.}\ \bibnamefont {Stacey}}}
  (\bibinfo {year} {2017}),\ \href@noop {} {\bibfield  {journal} {\bibinfo
  {journal} {Axioms}\ }\textbf {\bibinfo {volume} {6}},\ \bibinfo {pages}
  {21}}\BibitemShut {NoStop}%
\bibitem [{\citenamefont {Gour}\ and\ \citenamefont {Kalev}(2014)}]{GSIC-POVM}%
  \BibitemOpen
  \bibfield  {author} {\bibinfo {author} {\bibnamefont {Gour}, \bibfnamefont
  {G.}}, \ and\ \bibinfo {author} {\bibfnamefont {A.}~\bibnamefont {Kalev}}}
  (\bibinfo {year} {2014}),\ \href@noop {} {\bibfield  {journal} {\bibinfo
  {journal} {J. Phys. A: Math. Theor.}\ }\textbf {\bibinfo {volume} {47}},\
  \bibinfo {pages} {335302}}\BibitemShut {NoStop}%
\bibitem [{\citenamefont {Holevo}(2001)}]{measurement-general}%
  \BibitemOpen
  \bibfield  {author} {\bibinfo {author} {\bibnamefont {Holevo}, \bibfnamefont
  {A.~S.}}} (\bibinfo {year} {2001}),\ \href@noop {} {\emph {\bibinfo {title}
  {Statistical Structure of Quantum Theory}}}\ (\bibinfo  {publisher}
  {Springer},\ \bibinfo {address} {Berlin})\BibitemShut {NoStop}%
\bibitem [{\citenamefont {Horodecki}\ \emph {et~al.}(2022)\citenamefont
  {Horodecki}, \citenamefont {Rudnicki},\ and\ \citenamefont
  {Zyczkowski}}]{FIVE-OPEN}%
  \BibitemOpen
  \bibfield  {author} {\bibinfo {author} {\bibnamefont {Horodecki},
  \bibfnamefont {P.}}, \bibinfo {author} {\bibfnamefont {L.}~\bibnamefont
  {Rudnicki}}, \ and\ \bibinfo {author} {\bibfnamefont {K.}~\bibnamefont
  {Zyczkowski}}} (\bibinfo {year} {2022}),\ \href@noop {} {\bibfield  {journal}
  {\bibinfo  {journal} {PRX Quantum}\ }\textbf {\bibinfo {volume} {3}},\
  \bibinfo {pages} {010101}}\BibitemShut {NoStop}%
\bibitem [{\citenamefont {Kalev}\ and\ \citenamefont {Gour}(2014)}]{MUM}%
  \BibitemOpen
  \bibfield  {author} {\bibinfo {author} {\bibnamefont {Kalev}, \bibfnamefont
  {A.}}, \ and\ \bibinfo {author} {\bibfnamefont {G.}~\bibnamefont {Gour}}}
  (\bibinfo {year} {2014}),\ \href@noop {} {\bibfield  {journal} {\bibinfo
  {journal} {NJP}\ }\textbf {\bibinfo {volume} {16}},\ \bibinfo {pages}
  {053038}}\BibitemShut {NoStop}%
\bibitem [{\citenamefont {Kimura}(2003)}]{Kimura}%
  \BibitemOpen
  \bibfield  {author} {\bibinfo {author} {\bibnamefont {Kimura}, \bibfnamefont
  {G.}}} (\bibinfo {year} {2003}),\ \href@noop {} {\bibfield  {journal}
  {\bibinfo  {journal} {Phys. Lett. A}\ }\textbf {\bibinfo {volume}
  {314.5-6}},\ \bibinfo {pages} {339}}\BibitemShut {NoStop}%
\bibitem [{\citenamefont {Kimura}\ and\ \citenamefont
  {Kossakowski}(2005)}]{Kossakowski}%
  \BibitemOpen
  \bibfield  {author} {\bibinfo {author} {\bibnamefont {Kimura}, \bibfnamefont
  {G.}}, \ and\ \bibinfo {author} {\bibfnamefont {A.}~\bibnamefont
  {Kossakowski}}} (\bibinfo {year} {2005}),\ \href@noop {} {\bibfield
  {journal} {\bibinfo  {journal} {Open Sys. and Information Dyn.}\ }\textbf
  {\bibinfo {volume} {12}},\ \bibinfo {pages} {207}}\BibitemShut {NoStop}%
\bibitem [{\citenamefont {Rastegin}(2014)}]{SIC-POVM2}%
  \BibitemOpen
  \bibfield  {author} {\bibinfo {author} {\bibnamefont {Rastegin},
  \bibfnamefont {A.~E.}}} (\bibinfo {year} {2014}),\ \href@noop {} {\bibfield
  {journal} {\bibinfo  {journal} {Phys.Scr.}\ }\textbf {\bibinfo {volume}
  {89}},\ \bibinfo {pages} {085101}}\BibitemShut {NoStop}%
\bibitem [{\citenamefont {Renes}\ \emph {et~al.}(2004)\citenamefont {Renes},
  \citenamefont {Blume-Kohut}, \citenamefont {Scott},\ and\ \citenamefont
  {Caves}}]{SIC-POVM1}%
  \BibitemOpen
  \bibfield  {author} {\bibinfo {author} {\bibnamefont {Renes}, \bibfnamefont
  {J.~M.}}, \bibinfo {author} {\bibfnamefont {R.}~\bibnamefont {Blume-Kohut}},
  \bibinfo {author} {\bibfnamefont {A.~J.}\ \bibnamefont {Scott}}, \ and\
  \bibinfo {author} {\bibfnamefont {C.~M.}\ \bibnamefont {Caves}}} (\bibinfo
  {year} {2004}),\ \href@noop {} {\bibfield  {journal} {\bibinfo  {journal} {J
  Math. Phys.}\ }\textbf {\bibinfo {volume} {45}},\ \bibinfo {pages}
  {2171}}\BibitemShut {NoStop}%
\bibitem [{\citenamefont {Schumacher}\ and\ \citenamefont
  {Alber}(2023{\natexlab{a}})}]{Schumacher-Alber-PRA}%
  \BibitemOpen
  \bibfield  {author} {\bibinfo {author} {\bibnamefont {Schumacher},
  \bibfnamefont {M.}}, \ and\ \bibinfo {author} {\bibfnamefont
  {G.}~\bibnamefont {Alber}}} (\bibinfo {year} {2023}{\natexlab{a}}),\
  \href@noop {} {\bibinfo  {journal} {Phys.Rev. A}\ ,\ \bibinfo {pages} {in
  print}}\BibitemShut {NoStop}%
\bibitem [{\citenamefont {Schumacher}\ and\ \citenamefont
  {Alber}(2023{\natexlab{b}})}]{Schumacher-Alber-PhysScr}%
  \BibitemOpen
\bibfield  {journal} {  }\bibfield  {author} {\bibinfo {author} {\bibnamefont
  {Schumacher}, \bibfnamefont {M.}}, \ and\ \bibinfo {author} {\bibfnamefont
  {G.}~\bibnamefont {Alber}}} (\bibinfo {year} {2023}{\natexlab{b}}),\
  \href@noop {} {\bibfield  {journal} {\bibinfo  {journal} {Phys.Scr.}\
  }\textbf {\bibinfo {volume} {98}},\ \bibinfo {pages} {115234}}\BibitemShut
  {NoStop}%
\bibitem [{\citenamefont {Siudzinska}(2022)}]{NMPOVM}%
  \BibitemOpen
  \bibfield  {author} {\bibinfo {author} {\bibnamefont {Siudzinska},
  \bibfnamefont {K.}}} (\bibinfo {year} {2022}),\ \href@noop {} {\bibfield
  {journal} {\bibinfo  {journal} {Phys. Rev. A}\ }\textbf {\bibinfo {volume}
  {105}},\ \bibinfo {pages} {042209}}\BibitemShut {NoStop}%
\bibitem [{\citenamefont {Wootters}\ and\ \citenamefont {Fields}(1989)}]{MUB}%
  \BibitemOpen
  \bibfield  {author} {\bibinfo {author} {\bibnamefont {Wootters},
  \bibfnamefont {W.~K.}}, \ and\ \bibinfo {author} {\bibfnamefont {B.~D.}\
  \bibnamefont {Fields}}} (\bibinfo {year} {1989}),\ \href@noop {} {\bibfield
  {journal} {\bibinfo  {journal} {Ann. Phys.}\ }\textbf {\bibinfo {volume}
  {191}},\ \bibinfo {pages} {363}}\BibitemShut {NoStop}%
\end{thebibliography}%
\end{document}